# The Snippets Taxonomy in Web Search Engines

Artur Strzelecki[1] [0000-0003-3487-0971] and Paulina Rutecka[1] [0000-0002-1609-9768]

[1] University of Economics in Katowice, Department of Informatics, Katowice 40-287, Poland
{artur.strzelecki}{paulina.rutecka}@ue.katowice.pl

**Abstract.** In this paper authors analyzed 50 000 keywords results collected from localized Polish Google search engine. We proposed a taxonomy for snippets displayed in search results as regular, rich, news, featured and entity types snippets. We observed some correlations between overlapping snippets in the same keywords. Results show that commercial keywords do not cause results having rich or entity types snippets, whereas keywords resulting with snippets are not commercial nature. We found that significant number of snippets are scholarly articles and rich cards carousel. We conclude our findings with conclusion and research limitations.

## 1   Introduction

Rich Snippets as a Google search engine element appeared on the Internet in 2012. It was a Google answer for changing how users asked a search engine. We can risk saying that the style of entering queries to the search engine evolved along with the generation. The X generation were the first global Internet users. They have formed queries in simple and password method. They have been trying to understand computers, learn how they work, assuming that the machine to which the question is being asked isn't intelligent. In response to this, webmasters prepare reflecting the form of the entered enquiry in the 1:1 relationship.

As the effect, they made difficult to read and understand content with low substantive value. Perfect fitting was the sole aim of these contents oneself into factors in the ranking of search engines. In 2005 - 2010 users have used search engines in the same way that they have used other software. They have tried to learn software, read the user manuals to use it efficiently. In accordance with it, the system of notation of enquiries introduced to the search engine arose collected and at present available in the table summing up types of fitting the keyword.

Google constantly optimizes the way the search engine works. The purpose of this is to make a valuable search engine results pages with interesting and highly reliable content. The search engine of Google was launched in 1997 and in the last 22 years, it elaborated mechanism concerning fitting moved closer more and more. It recognized next variants of the enquiry: the variety, synonyms or mistakes of the spelling.

The revolution in search engine have started with a changing generation of computer users. The computer is now a companion of any person from Y generation who grew up with global access to the Internet global. A computer has become not just a working tool but a communication tool. It allows to access knowledge and entertainment. The Y generation doesn't try to learn how a computer works. They took it for granted and they do not attach special importance to learning this (except specialist skills).

Queries entered into the search engine have also become more natural and computer have become to being a partner in discussion. Queries become very similar to the question of which person can ask one another, preserving the syntax characteristic for questions, starts with adverbial - who, when, why, how etc.

The insertion of elements AI to search engine allowed for proper recognition of these types of queries and the evolution of the results display system in the search engine. In relation to change of the type of enquiries, increasing the number of vocal enquiries, leading into use the vocal assistant Google, it is possible to state that different Rich Snippets kinds are a natural reply to the demand of the market.

Establishing the research material of what type and the kind based on conducted analysis is a goal of the present article keywords cause the Snippets appearance in the search engine. It will enable further research above the strategy of building the plot up to get this position in search engine and the assessment of the impact of these results to the value of websites from which he is being downloaded content. We propose following research questions:

1. Is search engine evolving into human oriented system?
2. How search engine answers to specific questions?

The aim of this study is to retrieve information, conduct analysis and draw contribution on search engines rich results. Added value of this study is, that based on real search data for 50k keywords along with displayed snippet results, authors proposed several observations of current rich results appearing in search engine. Rising importance of search features like scholarly articles and direct answer (also known as featured snippet) was noticed.

The paper is organized as follows. Section 2 contains a review of the relevant literature on the topic. Section 3 includes the concept of the snippets taxonomy, while section 4 presents the data and quantitative results. In section 5, the authors highlight the contribution of the research, discuss its limitations and, finally, draw conclusions about the results and propose possible future research avenues.

## 2 Literature review

Snippets in search engines can be considered in five areas. The first area is regular snippets generated for regular, organic results. Four years regular snippets were two lines of description presented below the title and url of displaying results [1]. Recently we can observe some tests of increasing its length either on desktop version or on mobile devices [2, 3]. Scientific interest in regular snippet is mainly whether they are enough informative for readers or not enough [4]. Some tests are done on different age groups to see how these regular snippets are perceived [5].

The second area is rich snippets created based on structured data [6]. Search engines like Google, Microsoft (Bing), Yahoo and Yandex founded schema.org and are able to recognize structured data provided in RDFa, Microdata or JSON [7]. Rich snippets based on structured data are added to regular snippets [8]. Search engines show additional data about product availability, price and condition, recipes, reviews, jobs, music, video and others, included in schema.org. Rich snippets appears to become a more important variable, especially when examining bottom-ranked results [9].

Third area is snippets generated in Google News. These snippets are created completely automatically [10]. These snippets are considered by news publishers in different ways. Recently in Spain or Germany Google news was restricted, cause displaying snippets of news releases violates copyrights of news publishers [11]. To solve this possible violation a plan for ancillary copyright is proposed, by creating original snippets [12].

The fourth area is featured snippets. This is one of a recent snippet type. The search engine extracts pieces of information from web pages and presents it in a box, above organic results along with a source url. Google programmatically determines that a page contains a likely answer to the user's question and displays the result as a featured snippet. The other working name for this snippet is a direct answer or answer box. Direct answer supposed to deliver answers for queries, without need to visit the result presented in search engine [13]. This snippet can be presented in several different forms like paragraph [14], table [15] and ordered or unordered list.

The fifth area is entity types. Entity types are known in Google as Knowledge Graph introduced in 2012 year and in Bing are known as Satori introduced in the same year [16]. These entities are constructed object and concepts, including people, places, books, movies, events, arts, science, etc. Creating and maintaining these entity databases is considered as an important responsibility for search engines [17]. Search engines can create objects displayed in search results and also they remove results because of the variety of reasons [18].

## 3    Snippets taxonomy

The authors collected data for analysis using Senuto. Senuto is an online service which collects data from Google search engine. Senuto has a database of 20 million keywords. Each keyword is at least once in a month entered to Polish localized Google search engine and a list of top 50 results is returned. Senuto checks what rich and features snippets appear next to your keywords in Google search. A dataset from senator was acquired in May 2018. The dataset contains a list of 50000 keywords and their metrics. The dataset was limited only to keywords which in results shows not only ten blue links, but also have other rich and feature snippets, displayed above and on the right side in Google's search engine results page. Basic metrics for this keyword dataset are: cost per click (cpc), number of words, the average number of monthly searches in a year, features of keyword, average number of monthly searches in each month.

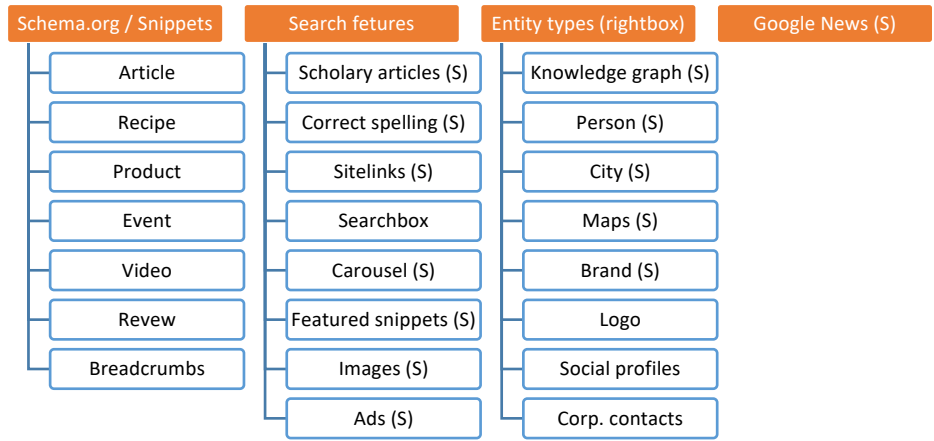

**Fig. 1.** Snippets taxonomy in Google search engine

Cost per click is estimated price per one click if this keyword would be used in sponsored search results. Number of words defines how long is the keyword. Average number of monthly searches is the number of how many times this keyword was entered into Google search. This number is limited to language. Cost per click and average number of searches is imported to Senuto from Google Planner. Google Planner is Google's tool, which shows metrics for keywords used in sponsored search results.

The most interesting aspect of this keyword dataset is that it contains keyword which cause displaying rich and feature snippets along with search results. Senuto distinguishes between 14 different rich and feature snippets. These 14 snippets are: ads (formerly AdWords), scholarly articles, correct spelling and grammar, Google news, knowledge graph, carousel, person, city, site links, maps, direct answer, right box, brand query and images.

Google Ads are results displayed in search engine results page which come from an advertising platform [19]. Scholarly articles is a featured snippet which contains around 3 results from Google Scholar together with author and number of citations [20]. Correct spelling and grammar is a snippet which suggests correct spelling and grammar form of provided query [21]. Google News aggregates news articles published in online newspapers and portals. Google News displays automatically results as a snippet together with image for results in a country, where Google News is available [22]. Knowledge Graph is a notion introduced to Google results in 2012. This feature is designed to sort and display known fact, places and persons [23].

The carousel is a graphical form to display similar results in one row above regular results. This placement is also called as knowledge card [24]. Carousel / knowledge card displays results in a structured order. These results are persons or cities. A query containing name and surname of a person which is known or popular artists (e.g. writer or actor) cause results as a set of work by this artist. Similar results looking as a carousel are presented for queries containing the names of cities.

Site links are results displayed only together with the first result mentioned. Site links are extending the first result by providing additional snippets and are only appearing when query is similar or the same as domain name appearing on first place in the ranking. The map is displayed for queries containing the name of known place which has a physical address. Direct answer is feature snippet containing a snippet with extracted answer for the query [25]. The direct answer is a box and usually contains a piece of text in the form of paragraph, table, ordered list or unordered list. Right box is known for displaying knowledge graph or a map [26]. There are types of queries which cause displaying results in right box, e.g. name of the book and author. In this case right box contains name of author, year of publishing and cover of the book. Brand query usually contains brand name and cause displaying in right box additional information about the brand. Images are displayed on result pages as one row, containing several images connected with a query.

## 4 Data and results

### 4.1 Data

The authors summarized the results in following tables. Table 1 presents the frequency of occurrence of snippet depending on the length of the keyword. Most keywords in the analyzed data set are 2 or 3 word-long words. Less popular, but still a large group are 4 or 5 word-long words.

**Table 1.** Keywords with specific number of words in every types of snippets.

| Snippet | Number of words | | | | | | | | | |
|---|---|---|---|---|---|---|---|---|---|---|
| | 1 | 2 | 3 | 4 | 5 | 6 | 7 | 8 | 9 | 10 |
| Ads | 0 | 9 | 39 | 14 | 3 | 0 | 0 | 0 | 0 | 0 |
| Scholarly articles | 29 | 307 | 9895 | 3488 | 1022 | 327 | 152 | 62 | 29 | 16 |
| Correct spell. and grammar | 4 | 147 | 311 | 53 | 6 | 3 | 1 | 0 | 0 | 0 |
| Google news | 0 | 0 | 1 | 0 | 0 | 0 | 0 | 0 | 0 | 0 |
| Carousel | 32 | 10255 | 17568 | 5226 | 1159 | 373 | 71 | 10 | 0 | 1 |
| Knowledge graph | 9 | 5131 | 12284 | 3312 | 837 | 361 | 97 | 28 | 9 | 4 |
| Person | 0 | 77 | 2346 | 504 | 117 | 62 | 16 | 4 | 0 | 0 |
| Site links | 6 | 250 | 518 | 199 | 54 | 11 | 3 | 0 | 1 | 0 |
| Maps | 0 | 750 | 2153 | 504 | 54 | 8 | 1 | 0 | 0 | 0 |
| City | 0 | 276 | 1225 | 315 | 33 | 10 | 0 | 0 | 0 | 0 |
| Direct answer | 5 | 5110 | 11921 | 3206 | 778 | 328 | 78 | 21 | 4 | 3 |
| Right box | 9 | 5131 | 12284 | 3312 | 837 | 361 | 97 | 28 | 9 | 4 |

| | | | | | | | | | |
|---|---|---|---|---|---|---|---|---|---|
| Brand query | 4 | 861 | 3180 | 1489 | 434 | 171 | 52 | 23 | 12 | 8 |
| Images | 45 | 5767 | 11734 | 3654 | 884 | 224 | 76 | 24 | 7 | 4 |

Table 2 presents correlations between snippets. Snippets have been divided into two parts. The first part contains most popular snippets. Second part contains snippets occurring less frequently, mostly together with another type of snippet. The second part of snippets is a peculiar group of answers for user's query, which appears in combination with the first set of snippets as a response to particular question containing eg. person, city, brand query.

**Table 2.** Correlations between snippets.

| | Person | Brand query | Images | City | Site links | Correct spelling and grammar | Google News |
|---|---|---|---|---|---|---|---|
| Ads | 9 | 2 | 25 | 1 | 1 | 0 | 0 |
| Scholarly articles | 3117 | 445 | 6560 | 72 | 91 | 81 | 1 |
| Carousel | 9 | 5789 | 15878 | 1787 | 951 | 444 | 0 |
| Knowledge graph | 0 | 3631 | 6829 | 1562 | 843 | 275 | 0 |
| Maps | 0 | 826 | 989 | 253 | 167 | 10 | 0 |
| Direct answer | 0 | 3637 | 6457 | 1562 | 843 | 275 | 0 |
| Right box | 0 | 3631 | 6829 | 1562 | 843 | 275 | 0 |

Table 3 presents a summary of the number of snippet instances and percentage of snippet instances. Table also shows average number of monthly searches for keywords that display snippet and median search volume.

**Table 3.** Summary of impressions and searches for keywords that display snippet.

| Snippet | Number of occurrences | % of dataset | Avg. number of monthly searches | Median monthly searches |
|---|---|---|---|---|
| Carousel | 34695 | 69,39% | 248 | 20 |
| Images | 22419 | 44,84% | 112 | 10 |
| Knowledge graph | 22072 | 44,14% | 108 | 20 |
| Right box | 22072 | 44,14% | 183 | 10 |
| Direct answer | 21454 | 42,91% | 183 | 10 |
| Scholarly articles | 15327 | 30,65% | 35 | 10 |
| Brand query | 6234 | 12,47% | 183 | 10 |
| Maps | 3470 | 6,94% | 135 | 30 |
| Person | 3126 | 6,25% | 83 | 20 |
| City | 1859 | 3,72% | 254 | 50 |
| Site links | 1042 | 2,08% | 2765 | 20 |
| Correct spelling and grammar | 525 | 1,05% | 67 | 10 |
| Ads | 65 | 0,13% | 90 | 20 |
| Google news | 1 | 0,00% | 1000 | 1000 |

**Google Ads.** The Google Ads snippet appeared merely 65 times as shown in Table 3. It is only 0,13% of analyzed keywords. The average monthly number of searches for all keywords where the Ads snippet has appeared is 90. The median of the monthly number of searches is 20. All keywords with Ads snippet were questions built out of 2 words (9 results), 3 words (39 results), 4 words (14 results) or 5 words (3 results) in the phrase as confirmed with data in Table 2. Correlations with other snippets were presented in Table 2. Ads snippet was displayed with name 9 times, with a brand query - 2 times, with images - 25 times, with city and site links - 1 time. Google News and correct spelling and grammar were not displayed.

**Scholarly articles.** Scholarly article snippets appeared 15327 times as shown in Table 3. For more than 30% of analyzed keywords search results were found in Google Scholar articles index and the snippet suggested by Google led the user to scientific papers. The data in table 1 shows the vast majority of keywords which have the snippet with scholarly articles as the result to user's query are long tail keywords. Most keywords have 3 words (9895 results), 4 words (3488 results) or 5 words (1022 results) in the phrase. The others have 1 word (29 results), 2 words (307 results), 6 words (327 results), 7 words (152 results), 8 words (62 results), 9 words (29 results), 10 words (16 results).

This means that user's query which causes the appearance of the snippet of the scholarly article are very exact due to the fact that users are looking for specific information. The analysis of individual words indicates that the majority of queries displaying this type of snippets concerns the field of exact and natural sciences ex.: physics, chemistry, medicine, IT. Table 2 presents correlations between keywords with scholarly articles and the other snippets. Scholarly articles were displayed with name (3117 times), brand query (445 times), images (6560 times), city (72 times), site links (91 times), Google news (1 time), correct spelling and grammar (81 times).

**Rich card carousel.** It is one of the most frequently showed snippet during keywords analysis in the research conducted by the authors. It has appeared for 69,39% of keywords that is, for 34695 records what Table 3 shows. Rich card carousel presents answer for user queries most often in a graphic form. In this type of snippets, the query has more than one answer and it is a list of possible answers in a graphic form of a carousel. The data in Table 1 indicates that most keywords that cause carousel snippet are phrases with 2 (10255 results) or 3 words (17568 results). They are rarely words 4 (5226 results), 5 (1159 results) and 6 (373 results) expressive. Keywords with a different number of words very rarely cause the occurrence of carousel snippets. Table 2 shows the correlation with others snippets and in this case. 45.76% (15878 results) of keywords with carousel have images at the same time. This shows the close connection of the carousel with the pictures. In second place in terms of the number of occurrences, there is a correlation between carousels and brand query (5789 results). Carousels also appear together with City (1787 results), Site Links (951 results) and Correct spelling and grammar (444 results).

**Knowledge Graph.** Knowledge Graph appeared for 44,14% of analyzed keywords (22072 results) independently or along with other snippets depending on the query construction what is show in Table 2. It occurs for such queries, that answer to which may be clearly defined as eg. first and last name, the name of the city or village. Other snippets appeared with Knowledge Graph are: brand query (3631 results), images (6829 results), city (1562 results), site links (843 results), correct spelling and grammar (275 results). Person snippet does not appear due to the frequent occurrence of a person inside the Knowledge Graph itself.

As shown in Table 1 for this type of snippets, 3 keywords are dominant (12284 results). The occurrence of 2 (5131 results) or 4 (3312 results) word-long words is also popular. Knowledge graph appeared for keywords with any number of words.

**Other snippets.** During the research authors also had analyzed other kinds of snippets such as:

- Name - appeared in 3126 analyzed records (6,43%)
- City - appeared in 1859 analyzed records (3,72%)
- Image - appeared in 22419 analyzed records (44,83%)
- Brand word - appeared in 6234 analyzed records (12,47%)
- Maps - appeared in 3470 analyzed records (6,94%)
- Sitelinks - appeared in 1042 analyzed records (2,08%)

**Additional indicators.** There were additional indicators in the set of data analyzed by the authors like CPC, number of words, the average monthly number of searches. These indicators were found to be of minor importance. Type of word and grammatical construction are, however, important.

A different border values in the data like CPC from 0.00 to 44.44, number of words from 1 to 10 or the average monthly number of searches from 10 to 2740000 indicate that there is no impact on the appearance of Snippets depending on these factors.

### 4.2 Results

The analysis of data clearly shows a dynamic growth and evolution of snippets in Google search engine. The types of snippets depend on the form of the question being asked, the keywords appearing (eg. games, movies for rich card carousel) or grammatical construction of query (eg. question form for featured snippets). The observations confirm that the development of the search engine is directed towards voice queries [27] and the user's dialogue with the search engine as an intelligent bot intended to provide specific answers. For most of the keywords, there is more than one type of snippets. The form of the answer given in snippets is short and shall be word or picture based. It encourages the user to read more information about the topic, which confirms the nesting of related headwords and interesting facts in the Knowledge Graph and links to the source page in Rich Answers.

A presentation for over 30% of keywords with answers containing references to scientific publications and target addresses of pages in Rich Answers, which lead to expert pages, confirms that Google in natural, non-advertising search results focuses on the reliability and highest quality of published content. This thesis is confirmed mainly by the results for the medical industry - referring to scientific articles. Google has also introduced an extensive list of medical-related keywords (including chemicals) for which advertising is prohibited. Snippets published on Google are user-friendly on mobile devices and are designed to be useful to users of voice search and chat with the Google Assistant.

## 5 Conclusion and Discussion

### 5.1 Discussion

In this paper, we presented an analysis of the set of data that causes Rich Snippets to appear in the search engine. The findings of our study indicate that the Google search engine is being developed in the direction of displaying the query response from the search results page. Google does not discriminate blue links, but makes the valuable site stand out. We collected data for 50000 keywords triggering in search results different types of snippets.

### 5.2 Contribution

Snippets content come from only reliable websites. The scale of the phenomenon (more than 30% of the keywords contains Snippets in the form of scholarly articles) confirms that Google is to improve its algorithms, trying to get the content of the highest quality distributed to the user from the most reliable source. Academic search behavior can be different from the web search behavior due to different types of contents, search goals and users [28], however placing results from scholarly articles is more and more often.

This paper is a first attempt to analyze the keywords which resulted in rich snippets in Polish localized Google search engine. Collected data reveal, that for 50k keyword rich snippets appear above organic results. The authors did analyze correlations between overlapping snippets. Correlations show that rich snippets are commercially independent. They usually do not appear for commercial keywords. Rich snippets appear with equal frequency for keywords with low CPC and for keywords with very high CPC. Estimated cost per click is not a defining factor defining the display of any type of Rich Snippet [29]. The keywords analysis shows that the keywords appearing in Google Ads have no influence on snippets appearance. Transactional [30] nature of the query is irrelevant to the appearance of snippets. Most of the keywords with active snippets do not cause displaying ads. Similarly, keywords displaying ads do not have snippets.

Google encourages users to use Rich Snippets by introducing an attractive visual form like in the Rich card carousel case. The image tiled display format, scrolled horizontally, is very mobile-user friendly and allows to present a large amount of information. It concentrates the user's attention, directing by just one click, to websites suggested by Google.

Rich card carousel applies for every query where the answer requires a list ex. titles of games or films, dog breeds or city districts. When the user uses the Google Assistant the result will be returned in the chat bubble or read by the voice assistant.

The Knowledge Graph is also a confirmation of the thesis regarding the credibility of websites used by Google to create Rich Snippet. These snippets in a short and concise way (2-3 sentence) answer for the user question. They also contain many links to subsequent searches that return results with different types of Rich Snippet. Knowledge Graph often appears in the company of a carousel, when it is necessary to present results in the list form.

Our results show that search engine results are more and more adapted to way users are asking questions and the answer is presented directly from results. This kind of solution belongs to human oriented systems.

### 5.3 Limitation

The limitation of our research was the fact of having a set of data concerning only the Polish language and only within 50,000 keywords. All data concern the Google search engine, which is dominant in Poland, but we realize that some types of Rich snippets can be observed in other search engines. The factors conditioning the appearance of specific types of Rich Snippets may be different in various search engines. Due to the lack of data, we did not analyze why a particular snippet appeared but only its type.

### 5.4 Future research

We acknowledge that Google strives to become the most reliable and user-friendly search engine and the snippet richness appears to become a more important variable, especially when examining bottom-ranked results [9]. Further testing will be conducted to investigate the factors affecting the display of results from specific websites in the snippets area. Also, further tests will be interesting for other languages.